# Theoretical estimate of the probability for macromolecule formation

Pramod Kumar Mishra
Email: pkmishrabhu@gmail.com
*Department of Physics, DSB Campus, Kumaun University, Nainital (Uttarakhand) India-263002*

**Abstract:** We estimate the probability regarding polymerization of a macromolecule which is made of distinct monomers; and there are different values of the fugacity for the addition of the monomers in the chain to form an infinitely long linear macromolecule of distinct monomers. The lattice model of the random walk has been used to mimic the conformations of an ideal chain in two and three dimensions. It has been shown through analytical estimates that the flexible macromolecules may be easily formed than the stiff macromolecules in two and three dimensions. In the case of stiff chain, the ratio of the critical value of monomer fugacity is nothing but the log-log ratio of the Boltzmann's weight corresponding to the monomers affinity corresponding to its conjugate monomer pair; and it is due to a fact that the stiff chain has small value of the entropy.

1. Introduction

There are four major types of the Biological macromolecules which are known to form various living organisms. These macromolecules are *Lipids*, *Carbohydrates*, *Nucleic Acids* and the *Proteins* [1]. The Lipids are non polymeric macromolecules. These Biological macromolecules are the backbone of the living creatures and the structures of these molecules are formed by distinct monomers and therefore understanding the possibility of polymerization of such macromolecules may be useful. There are artificial macromolecules which are known to be useful for various applications [2-3]. The macromolecules are formed by the process of dehydration; and the solute molecules which are in the solutions and acts as the monomers and these monomers may combine to form long macromolecules and these Bio-molecules are evolved in the complex manner [1-4].

We consider structure of a macromolecule in the form of random polymer chain which is organized in the two and three dimensions. The structure of the macromolecule has been realized using a square lattice and a cubic lattice to study polymerization of such structure of an infinitely long linear macromolecule in two and three dimensions respectively to mimic the formation of an infinitely long linear macromolecule made of four different repeat units or the monomers. The distinctions have been incorporated in the macromolecule in such a manner that one monomer may connect to another type of monomer with a given Boltzmann weight while addition of other type of monomers with third type of the monomers have different Boltzmann weight. Such means has been used to mimic conformations of a macromolecule which is made of different types of the monomers so that we have some theoretical estimates about the structural aspects of the complex macromolecules through over simplified mathematical model.

The manuscript is organized in the following manner; in the section two we define the random walk lattice model and we described methods for calculating the partition function of an infinitely long chain where the polymer chain is made of different type of the monomers.

We analyze results obtained through simplified artificial model of the linear chain to mimic conformations of the chain and also to calculate the estimate for the possibility of the formation of an infinitely long linear macromolecule in the section three. In the section four, we summarize results obtained and also we conclude the discussion in this section.

## 2. Model and Method

We consider a linear ideal polymer chain to mimic the conformations of a macromolecule in two and three dimensions. The conformations of an ideal polymer chain were enumerated using under lying lattices, i. e. a square and a cubic lattice for two and three dimensions, respectively. The lattice models were widely used to study the thermo-dynamical behaviour of linear polymer molecules [5-13] in the thermodynamic limit; and we have also relied to lattice model of an ideal polymer chain to analyze the possibility of the macromolecule formation. The generating function technique is used to enumerate conformations of the polymer chain [5,8-11]; and all possible directions i. e. $\pm x$ and $\pm y$ directions on the square lattice were accessible to the walker for enumerating the chain conformations, while in the cubic lattice $\pm z$ were also permitted to the walker in addition to $\pm x$ and $\pm y$ directions while enumerating the conformations of the polymer chain in three dimensions.

The chain is made of four type of monomers *a*, *b*, *c* and *d;* where bonding between *a* & *b* monomers leads release of an extra energy equal to $2*E_s$ and similarly bonding in between *c* and *d* monomers leads release of an extra energy $3*E_s$. The first monomer of the chain is grafted at a point O, as shown in the figure no. 1, and a walk of four monomers is shown in the Figure no. 1 (i) and the Boltzmann weight of this walk is $g^5 k^4$ and the Boltzmann weight of the walk shown in the Figure no. 1 (ii) is $g^9 k^7 \omega^4$. The semi-flexible behaviour is incorporated in the model by adding the stiffness weight in the conformations of the chain.

Since, it is assumed that there are four different kinds of the monomers present in the solution, and these monomers are assumed to form the macromolecule by the process like dehydration. Any of these four type of the monomers may connect to remaining three type of the monomers and the Boltzmann weight corresponding to addition of the monomer is at least *g* step fugacity; however while *a* type monomers can connect to *b* type monomer with the Boltzmann weight $g\omega^2$ [where $\omega=\mathrm{Exp}(\beta E_s)$]; while $g\omega^3$ is the Boltzmann weight corresponding to connection of the *c* type monomers to the *d* type monomers. The Boltzmann weight $k[=\mathrm{Exp}(\beta E_b)]$ corresponds to bending energy 90 degree while 180 degree bending energy corresponds to the stiffness weight $k^2$ [12-13].

The grand canonical partition function of the polymer chain in general may be written as,

$$F(g,\omega,k) = \sum_{N=1}^{N\to\infty} \sum_{n=0}^{\infty} \sum_{N_B=0}^{N-1} g^N \omega^n k^{N_B} \qquad (1)$$

## 3. The Calculations and the Results

We use the generating function technique [5-14] to obtain the partition function of the polymer chain in the thermodynamic limit and the singularity of the partition function of the chain is used to find condition for the polymerization of an infinitely long chain. The admissible value of $\omega_c$ is shown in the table no. 1 and variation of the corresponding phase diagram is shown in the figure no.1. It is to be noted here that $\omega_c$ is the minimum weight required for the polymerization of an infinitely long chain of different monomers while $g_c$ is the weight for the formation of an infinitely long chain of identical monomers.

We obtained the values of the relevant parameters i. e. the average number of *a-b* monomer pairs and *c-d* monomer pairs using following equations,

$$<N_u> = \frac{\partial Log(F)}{\partial Log(u)} \qquad (2)$$

$$<N_v> = \frac{\partial Log(F)}{\partial Log(v)} \qquad (3)$$

Where $u = g*\omega^2$ and $v = g*\omega^3$, and the average length of the chain is obtained using following the well known equation,

$$<N_g> = \frac{\partial Log(F)}{\partial Log(g)} \qquad (4)$$

The fraction of *a-b* and *c-d* monomer pairs may be obtained using following relations,

$$n_a = \frac{<N_u>}{<N_g>}, \qquad n_b = \frac{<N_v>}{<N_g>} \quad \& \quad n_c = \frac{n_a}{n_b} \qquad (5)$$

### (a) Macromolecule in two dimensions

A monomer of the chain is assumed to polymerize in two dimensions and there are four types of the monomers forms an infinitely long chain. The grand canonical partition function (*F*) and average number of *a-b* and *c-d* monomer pairs may be written for two dimensional case in the following manner (here $g_c = 1/4$, [5-6]),

$$<N_u> = -\frac{e^{Eb+2Es}\left(1+e^{Eb}+3e^{2Eb}-e^{Eb+3Es}\right)^2}{\left(2+2e^{Eb}+6e^{2Eb}-e^{Eb+2Es}-e^{Eb+3Es}\right)\left(1+2e^{Eb}+e^{2Eb}-9e^{4Eb}+3e^{3(Eb+Es)}+3e^{3Eb+2Es}-e^{2Eb+5Es}\right)} \qquad (6)$$

$$<N_v> = \frac{e^{-Eb+3Es}\left(1+e^{Eb}+3e^{2Eb}-e^{Eb+2Es}\right)^2}{4\left(2+4e^{Eb}+14e^{2Eb}+12e^{3Eb}+18e^{4Eb}-2e^{2(Eb+Es)}-6e^{3(Eb+Es)}-2e^{Eb+2Es}-6e^{3Eb+2Es}-2e^{Eb+3Es}-2e^{2Eb+3Es}+e^{2Eb+4Es}+e^{2Eb+6Es}\right)} \qquad (7)$$

And the length of the chain may be written using following equation,

$$<N_g> = -\frac{4e^{2Eb}\left(2+4e^{Eb}+14e^{2Eb}+12e^{3Eb}+18e^{4Eb}-2e^{2(Eb+Es)}-6e^{3(Eb+Es)}-2e^{Eb+2Es}-6e^{3Eb+2Es}-2e^{Eb+3Es}-2e^{2Eb+3Es}+e^{2Eb+4Es}+e^{2Eb+6Es}\right)}{\left(2+2e^{Eb}+6e^{2Eb}-e^{Eb+2Es}-e^{Eb+3Es}\right)\left(1+2e^{Eb}+e^{2Eb}-9e^{4Eb}+3e^{3(Eb+Es)}+3e^{3Eb+2Es}-e^{2Eb+5Es}\right)} \qquad (8)$$

The fraction of *a-b* monomer pair and *c-d* monomer pair is written as,

$$n_a = \frac{e^{-\text{Eb}-2\text{Es}}(1+e^{\text{Eb}}+3e^{2\text{Eb}}-e^{\text{Eb}+3\text{Es}})^2}{4(2+4e^{\text{Eb}}+14e^{2\text{Eb}}+12e^{3\text{Eb}}+18e^{4\text{Eb}}-2e^{2(\text{Eb}+\text{Es})}-6e^{3(\text{Eb}+\text{Es})}-2e^{\text{Eb}+2\text{Es}}-6e^{3\text{Eb}+2\text{Es}}-2e^{\text{Eb}+3\text{Es}}-2e^{2\text{Eb}+3\text{Es}}+e^{2\text{Eb}+4\text{Es}}+e^{2\text{Eb}+6\text{Es}})}$$

(9)

$$n_b = \frac{e^{-\text{Eb}-3\text{Es}}(1+e^{\text{Eb}}+3e^{2\text{Eb}}-e^{\text{Eb}+2\text{Es}})^2}{4(2+4e^{\text{Eb}}+14e^{2\text{Eb}}+12e^{3\text{Eb}}+18e^{4\text{Eb}}-2e^{2(\text{Eb}+\text{Es})}-6e^{3(\text{Eb}+\text{Es})}-2e^{\text{Eb}+2\text{Es}}-6e^{3\text{Eb}+2\text{Es}}-2e^{\text{Eb}+3\text{Es}}-2e^{2\text{Eb}+3\text{Es}}+e^{2\text{Eb}+4\text{Es}}+e^{2\text{Eb}+6\text{Es}})}$$

(10)

While ratio of *a-b* pairs to that of *c-d* pairs is written as,

$$n_c = \frac{e^{-\text{Es}}(1+e^{\text{Eb}}+3e^{2\text{Eb}}-e^{\text{Eb}+3\text{Es}})^2}{(1+e^{\text{Eb}}+3e^{2\text{Eb}}-e^{\text{Eb}+2\text{Es}})^2}$$

(11)

We have shown variations of the parameters $n_a$, $n_b$ and $n_c$ in the figure no. (3) for the two dimensional model for the chosen values of energy of monomers pairing and the bending energy of the chain. It is shown that variation of $n_b$ is complex involved.

### (b) Macromolecule in three dimensions

We followed the method described in the above section 3 (a) regarding calculation of the relevant parameters of interest and plotted the variation of the parameters $n_a$, $n_b$ and $n_c$ in the figure no. (4) for the three dimensional model. We have also shown possible critical values of the monomer-monomer pairing energy for different stiffness weight of the chain in two and three dimensions in the table no. 1 for the sake comparison and the completeness.

### 4. Summary and Conclusions

We model the Gaussian polymer chain using square and cubic lattices to enumerate conformations of a polymer chain which is made of four different types of the monomers (*a*, *b*, *c* and *d*). The Gaussian chain is shown schematically in the two and three dimensions using Figure no. (1). The chain is polymerized in the thermodynamic limit provided $g_c=1/4$ for square and $g_c=1/6$ for the cubic lattices, respectively [5-6]. There are two additional critical values of the fugacity of an infinitely long chain which corresponds to divergence in the number density fluctuations of the *a-b* monomers pair and also *c*, *d* monomers pair. In the proposed model system, it is assumed that there are two hydrogen bonds in between *a, b* monomers while there are three hydrogen bonds in between *c, d* monomers. Therefore, we have chosen $u=\omega^2$ and $v=\omega^3$ and the powers on the Boltzmann weight $\omega$ is nothing but number of the hydrogen bonds in between these monomers i. e. (*a, b*) and (*c, d*) monomer pairs.

There are two values of $\omega$ which corresponds to singularity of the partition function and it has been found that the value of $Log[\omega_{c2}]/Log[\omega_{c1}]=1.5$ in the stiff chain limit for the two and three dimensional models. It is due to fact that chain has small entropy and $Log[v_c]/Log[u_c]=1.5$. However, we have shown $\omega_{c1}$ & $\omega_{c2}$ in the table no. 1 for different values of stiffness weight for two and three dimensional models.

An infinitely long chain is polymerized in two and three dimensions where the chain is made of *a-b* and *c-d* monomers pairs, as shown in the figure no. (2). Nature of variation of the relevant parameters *i. e. $n_a$*, *$n_b$* and *$n_c$* for the possible values of the monomers pairing energy ($E_s$) and the bending energy ($E_b$) is graphed in the figure no. (3) for the square lattice; and these parameters (*i. e. $n_a$, $n_b$* and *$n_c$*) are also shown in the figure no. (4) for the cubic lattice cases, respectively.

There are reports on single macromolecule [15-17], and we may infer little useful information regarding macromolecule using an oversimplified such models for the polymer chain. Where the chain is seen as the Gaussian chain and the conformations of the chain were enumerated using a square and a cubic lattice; the stiffness weight is used to mimic semi-flexible nature of the chain. The distinction in the monomers was incorporated using different values for the fugacity of different type of the monomers. It is shown in the figure no. (2) that monomer-monomer pairing energy increases with the bending energy of the chain, and *c-d* pair fraction and *a-b* pair fraction differ from each other and *a-b* monomers pair to *c-d* monomer pairs are also the function energy $E_s$ and $E_b$.

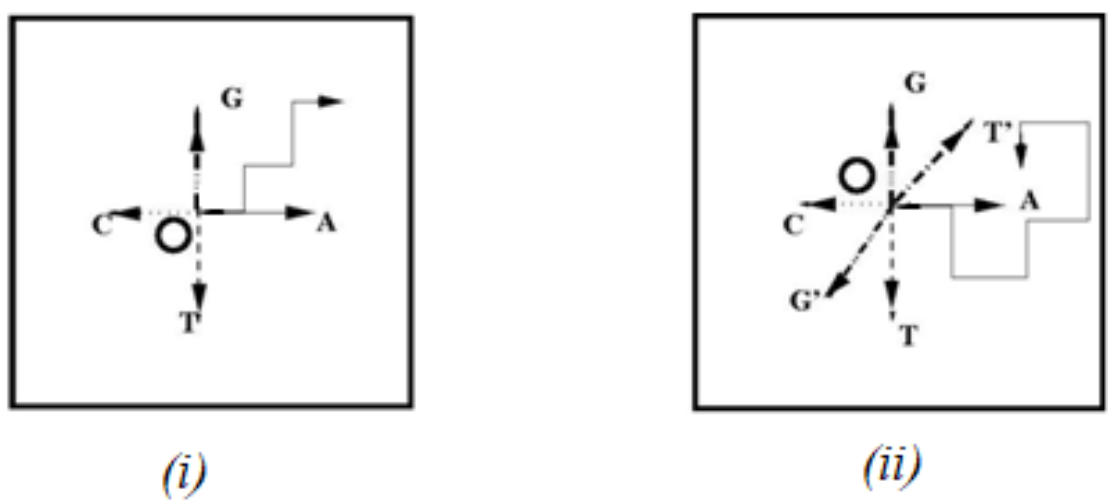

**Figure No. 1:** This figure (i) shows the components of the partition function of the Gaussian polymer chain in two dimensions, i. e. Figure No. 1 (i); as well as these components are shown in the three dimensions using the Figure no. 1 (ii). The terms *A*, *T*, *C* and *G* as shown in figure (i) shows the sum of the Boltzmann weight of all walks on a square lattice whose first step is along +*x*, -*y*, -*x* and +*y* directions, respectively. While in the figure no. 1 (ii), there are additional terms *T'* and *G'* those corresponds to the sum of the Boltzmann weights of all the walks whose first step is along +z and –z directions, respectively. The Boltzmann weight of the walk which is shown in Figure no. 1(i) is $g^5 k^4$ and the Boltzmann weight of the walk shown in the Figure no. 1 (ii) is $g^9 k^7 \omega^4$.

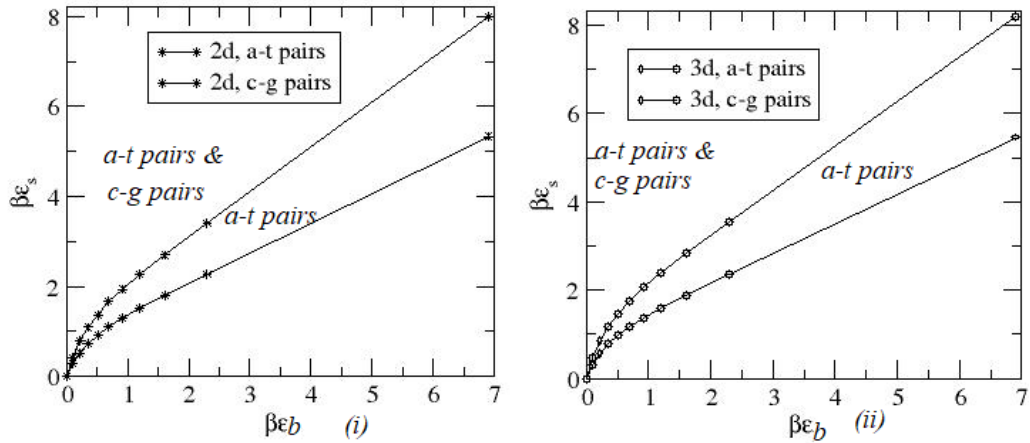

**Figure No. 2:** This figure shows the variation of the critical value of the monomers pairing energy ($E_s$) of an infinitely long chain for two and three dimensions. In the figure no. 2 (i), the variation of energy of the *a-b* monomers & *c-d* monomers affinity for admissible values of the bending energy ($E_b$) is shown for two dimensions; while in the figure no. 2 (ii) we have shown the variation of critical values of the energy for *a-b* monomers & *c-d* monomers affinity ($E_s$) for given values of bending energy ($E_b$) of the chain for three dimensions.

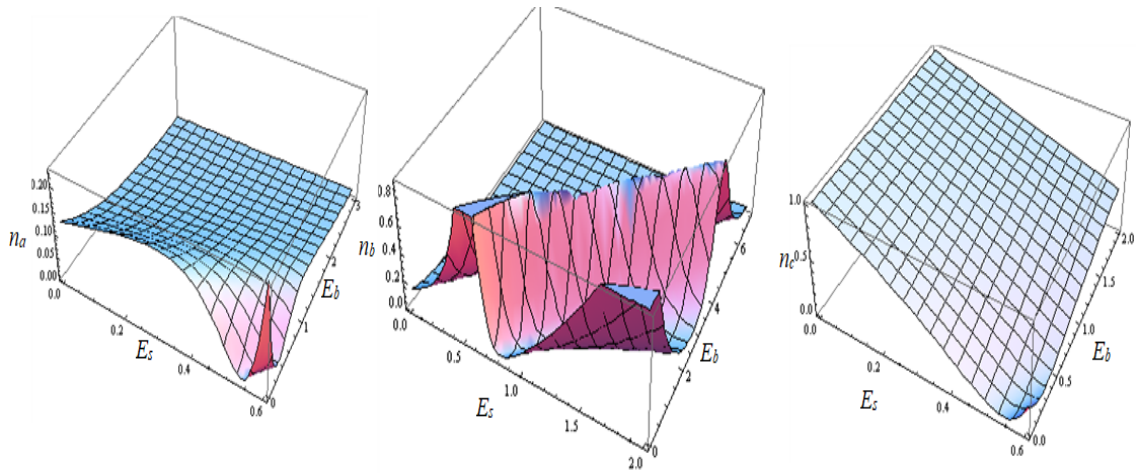

**Figure No. 3:** We have variation of the fractions $n_a$, $n_b$, and $n_c$ for two dimensional case is shown for the value of monomer-monomer pairing energy ($E_s$) and the bending energy ($E_b$) of the polymer chain. The energy values $E_s$ & $E_b$ are shown in the units of the inverse thermal energy ($\beta$), and for the sake of mathematical simplicity we have taken the value of $\beta=1$.

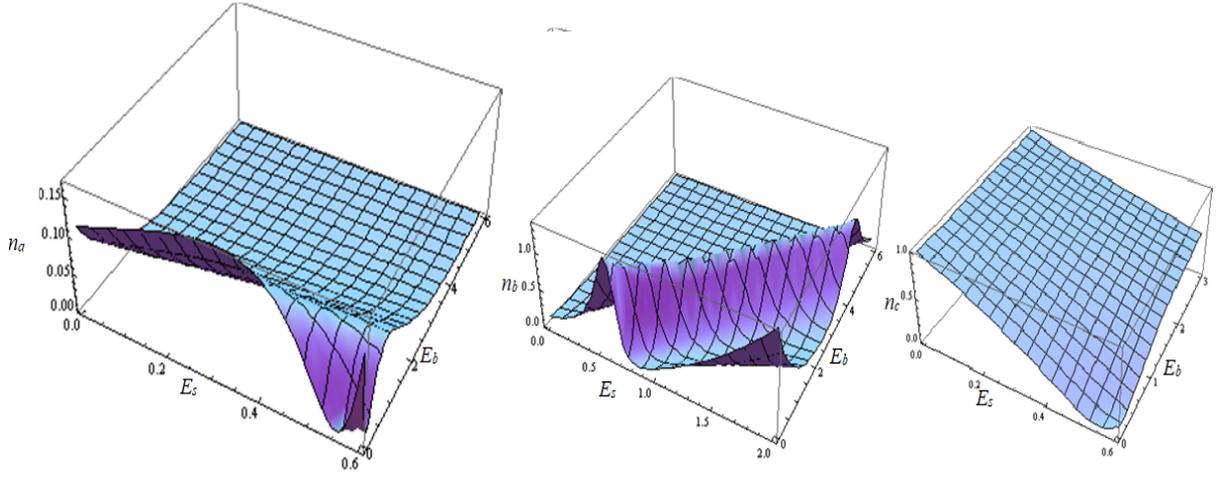

**Figure No. 4:** We have shown the variation of the relevant parameter $n_a$, $n_b$ and $n_c$ for the possible values of the monomer-monomer pairing energy ($E_s$) and the bending energy ($E_b$). These energy values were reported in the unit of β and the β=1 for the sake of mathematical simplicity in the proposed model. This figure corresponds to the case of an infinitely long chain which is made of four type of the monomers *a*, *b*, *c* and *d* and the chain is polymerised the three dimensions.

**Table No. 1:** We have shown the critical values of the pairing fugacity for the monomers of the chain to form an infinitely long macromolecule; the molecule is made of four different types of the monomers for two dimensional, and the macromolecule is made of six monomers in three dimensional model.

| K | $\omega_{c1}(2d)$ | $\omega_{c2}(2d)$ | $\omega_{c1}(3d)$ | $\omega_{c2}(3d)$ |
|---|---|---|---|---|
| 0.001 | 14.42 | 54.77 | 15.2336 | 59.4574 |
| .1 | 3.105 | 5.478 | 3.26836 | 5.91704 |
| .2 | 2.457 | 3.877 | 2.5765 | 4.16654 |
| .3 | 2.13 | 3.17 | 2.22621 | 3.39313 |
| .4 | 1.91 | 2.76 | 1.98845 | 2.93854 |
| .5 | 1.74 | 2.5 | 1.79985 | 2.63777 |
| .6 | 1.58 | 2.31 | 1.63429 | 2.42679 |
| .7 | 1.44 | 2.18 | 1.47887 | 2.27384 |
| .8 | 1.3 | 2.087 | 1.32599 | 2.16041 |
| .9 | 1.15 | 2.018 | 1.16926 | 2.07465 |
| .999 | 1.0016 | 1.9672 | 1.0018 | 2.00929 |